\documentclass[12pt]{article}
\newcommand{\be}{\begin{equation}}
\newcommand{\ee}{\end{equation}}

\newcommand{\bi}{\begin{itemize}}
\newcommand{\ei}{\end{itemize}}
\newcommand{\bea}{\begin{eqnarray}}
\newcommand{\eea}{\end{eqnarray}}
\newcommand{\ba}{\begin{array}}
\newcommand{\ea}{\end{array}}

\usepackage[left=2.50cm, right=2.50cm, top=2.50cm, bottom=2.50cm]{geometry}
\usepackage[utf8]{inputenc}
\usepackage{cancel}
\usepackage[english]{babel}
\usepackage{physics}
\usepackage{hyperref}
\usepackage{amsthm}
\usepackage{mathtools}
\usepackage{graphicx}
\usepackage{ulem}
\usepackage{changepage}
\usepackage{amssymb,amsmath}
\usepackage{verbatim}
\usepackage{empheq}
\usepackage{xcolor}
\usepackage{tikz}
\usepackage{float}
\usepackage{multirow}
\usepackage{multicol}
\usepackage{amsmath}
\usepackage{centernot}
\usepackage[hang,flushmargin]{footmisc} 
\usetikzlibrary{tikzmark}
\numberwithin{equation}{section}
\hypersetup{hidelinks}

\newlength{\bibitemsep}
\newlength{\bibparskip}
\let\oldthebibliography\thebibliography
\renewcommand\thebibliography[1]{%
  \oldthebibliography{#1}%
  \setlength{\parskip}{\bibitemsep}%
  \setlength{\itemsep}{\bibparskip}%
}
\begin{document}
\par
\bigskip
\Large
\noindent
{\bf 
Gauging fractons and linearized gravity\\
\bigskip
\par
\rm
\normalsize

\hrule

\vspace{1cm}

\large
\noindent
{\bf Erica Bertolini$^{1,2,a}$},
{\bf Alberto Blasi$^{1,b}$}, 
{\bf Andrea Damonte$^{2,c}$},\\
{\bf Nicola Maggiore$^{1,2,d}$}\\

\par
\small
\noindent$^1$ Dipartimento di Fisica, Universit\`a di Genova, Italy.
\smallskip

\noindent$^2$ Istituto Nazionale di Fisica Nucleare (I.N.F.N.) - Sezione di Genova, Italy.

\smallskip

\vspace{1cm}

\noindent
{\tt Abstract~:}
We consider the covariant gauge field theory of fractons, which describe a new type of quasiparticles exhibiting novel and nontrivial properties. In particular, we focus on the field theoretical peculiarities which characterize this theory, starting from the fact that,  if we accept the paradigm that quantum field theories are defined by their symmetries, fractons unavoidably come together with linearized gravity. The standard Faddeev-Popov procedure to gauge fix the theory leads to a scalar gauge condition, which has two important drawbacks: it is frozen in the Landau gauge and linearized gravity cannot be obtained as a limit. In this paper we adopt a tensorially alternative gauge fixing, which avoids both problems. In particular, this allows to show that important physical features, like the counting of the degrees of freedom, do not depend on a particular gauge choice, as expected. Moreover, the resulting gauge fixed theory contains both fractons and linearized gravity as a limit, differently from the standard scalar choice.

\vspace{.5cm}

\vspace{\fill}
\noindent{\tt Keywords:} 
Quantum Field Theory, Gauge Field Theory, Fractons, Linearized Gravity\\
{\tt Reference:} Symmetry {\bf 2023}, 15(4), 945; \url{https://doi.org/10.3390/sym15040945}.

\vspace{1cm}

\hrule
\noindent{\tt E-mail:
$^a$erica.bertolini@ge.infn.it,
$^b$alberto.blasi@ge.infn.it,
$^c$andrea@damonte.it,
$^d$nicola.maggiore@ge.infn.it.
}
\newpage

\section{Introduction}\label{sec:introduction}

In recent years there has been a surge of interest in a kind of exotic particles called ``fractons'', which have come to the forefront of modern condensed matter theory \cite{Prem:2017kxc,Pretko:2016kxt,Pretko:2016lgv,Pretko:2018jbi,Nandkishore:2018sel,Pretko:2020cko,Seiberg:2020wsg,Chamon:2004lew,Vijay:2015mka,Vijay:2016phm,Pretko:2017xar,Prem:2017qcp,Haah:2011drr,Bravyi:quantum,Shi:2017qdx,Slagle:2018kqf,Argurio:2021opr,pretkoscreening,Gromov:2018nbv, Gorantla:2022eem}. 
They represent a new class of quasiparticles, whose main property is that of being immobile in isolation, but may move by forming bound states. Fractons are found in a variety of physical settings, such as spin liquids \cite{Pretko:2016lgv} and elasticity theory \cite{Pretko:2017kvd,Pretko:2019omh}, and exhibit unusual phenomenology, such as gravitational physics and localization \cite{Xu:2006,Gu:2009jh,Xu:2010eg,Pretko:2017fbf}.
	Fractonic behaviours are to be connected with exotic global symmetries \cite{Seiberg:2020wsg,Vijay:2016phm,Gromov:2018nbv}. Systems with these symmetries challenge the common lore by which the low energy behaviour of every lattice system can be described by a continuum quantum field theory: these lattice constructions are usually not Lorentz invariant, and they present an unusual ground state degeneracy, infinite in the continuum limit. Significant efforts have been made to better understand their non-standard behaviours and to extend the known theoretical frameworks to include them \cite{Pretko:2018jbi,Pretko:2020cko,Seiberg:2020wsg,Gromov:2018nbv,Gorantla:2022eem,Blasi:2022mbl,Bertolini:2022ijb}. Our aim here is to study the theory of fractons from the field theoretical point of view where, again, peculiar and unusual features appear. In our approach, fractons are described by a gauge field theory involving a symmetric tensor field $h_{\mu\nu}(x)$ transforming as 
\be
\delta h_{\mu\nu} = \partial_\mu\partial_\nu \Phi\ ,
\label{sym}\ee
where $\Phi(x)$ is a scalar local gauge parameter. The transformation \eqref{sym} is a particular case of the infinitesimal diffeomorphisms
\be
\delta_{diff} h_{\mu\nu} = \partial_\mu\xi_\nu + \partial_\nu \xi_\mu\ ,
\label{diff}\ee
when the vector gauge parameter $\xi_{\mu}(x)$ is the derivative of a scalar field 
\be
\xi_{\mu}=\frac{1}{2}\partial_\mu\Phi\ .
\label{difftofract}\ee
The most general action $S_{inv}$ invariant under the fracton symmetry \eqref{sym} is the sum of two terms, one of which can be immediately identified as Linearized Gravity (LG) \cite{Carroll.book,Hinterbichler:2011tt}. This term is to be expected since the infinitesimal diffeomorphisms \eqref{diff}, which is the defining symmetry of LG, embeds the fracton symmetry \eqref{sym} through \eqref{difftofract}. The other term 
describes a theory with pure fractonic features \cite{Bertolini:2022ijb}: from this term one can recover all the properties characterizing the so called ``scalar charge theory'' of fractons \cite{Pretko:2016lgv}, including Maxwell-like equations of which the Gauss law is at the origin of the limited mobility property that defines fractons \cite{Prem:2017kxc,Pretko:2016kxt,Pretko:2016lgv,Nandkishore:2018sel,Pretko:2020cko}. This constraint on the motion is realized as a conservation equation, in the same way as in the standard Maxwell theory the Gauss law implies charge conservation. In fracton models, the conserved quantity is the dipole moment of the system, due to which, single, isolated, charges are to be immobile, and are thus identified as fractons. This inability to move is the main property shared by all fracton models, and ultimately can be of physical interest, for instance, in the development of quantum memories \cite{Haah:2011drr,Bravyi:quantum,Terhal,Ma:2017igk}. Since the transformation \eqref{sym} is a particular case of the more general infinitesimal diffeomorphism \eqref{diff}, which is a gauge transformation \cite{Carroll.book,Hinterbichler:2011tt,Blasi:2015lrg,Gambuti:2020onb}, 
the theory of the symmetric tensor field $h_{\mu\nu}(x)$ defined by the symmetry \eqref{sym} is a $gauge$ field theory as well. 
Evidence of this appears when trying to compute the propagator from the quadratic action $S_{inv}$, which, similar to the electromagnetic Maxwell theory, leads to a non invertible matrix. Therefore a gauge fixing term must be added. For any other gauge field theory, this procedure goes smoothly thanks to the Faddeev-Popov procedure \cite{Faddeev:1967fc}, which focuses on the gauge parameter, and in particular on its tensorial character. In \eqref{sym} the gauge parameter is a scalar, and this would require a scalar gauge fixing condition. The most general covariant one is
\be
\partial^\mu\partial^\nu h_{\mu\nu} + \kappa\partial^2h=0\ ,
\label{scalargaugecond}\ee
where $\kappa$ is a constant gauge fixing parameter. This is analogous to the covariant Lorentz condition for the vector field $A_\mu(x)$
\be
\partial^\mu A_\mu=0\ .
\label{lorentzgaugecond}\ee
The standard situation in gauge field theory is that of having a gauge field, represented by a $p$-tensor field, and a $(p-1)$-tensorial gauge parameter. This is the case of all known gauge theories. We already mentioned Maxwell theory, or its non-abelian counterpart, the Yang-Mills theory, but this is also true for higher rank theories, like the topological $BF$ theories in any dimensions \cite{Birmingham:1991ty,Bertolini:2022sao}. Linearized Gravity, described by a symmetric tensor field and by the symmetry \eqref{diff}, does not escape this rule. In this case, the commonly used gauge fixing condition is vectorial \cite{Hinterbichler:2011tt,Blasi:2017pkk,Gambuti:2021meo}
\be
\partial^\nu h_{\mu\nu} + \kappa\partial_\mu h=0\ .
\label{vectorgaugecond}\ee
Due to the presence of the parameter $\kappa$, the gauge fixing conditions \eqref{scalargaugecond} and \eqref{vectorgaugecond}  represent a class of covariant gauges. For instance, in \eqref{vectorgaugecond} the particular case $\kappa=-\frac{1}{2}$ corresponds to the ``harmonic'' gauge fixing \cite{Carroll.book,Gambuti:2020onb}. The theory we are considering here is, again, quite peculiar. Not only because, as we shall see, it displays a ``non-coupling'' constant, but also because it is defined by the gauge transformation \eqref{sym}, which associates a scalar gauge parameter to a rank-2 symmetric tensor field. Concerning the gauge fixing procedure, one might therefore look at both sides of \eqref{sym}, each of which opens a different path. Looking at the right hand side of \eqref{sym}, one sees a scalar gauge parameter and this leads to adopt the scalar gauge condition \eqref{scalargaugecond}. This standard way has been followed in \cite{Blasi:2022mbl}, where the propagators have been derived and the degrees of freedom have been studied. The scalar gauge condition \eqref{scalargaugecond} has two important drawbacks. The first is that the Landau gauge $\xi=0$ turns out to be mandatory. The theory seems not to be defined outside this gauge. Now, it is true that physical results should not depend on the gauge choice, but, still, being forced to a unique choice is unpleasant, and it would be much preferable to find all the physical results in a generic gauge and to show that they do not, indeed, depend on a particular choice. The second reason is that the scalar gauge condition \eqref{scalargaugecond} does not allow to reach the limit of pure LG, which necessarily needs the vector gauge condition \eqref{diff}. This results in a singularity both in the propagators of the theory and in the degrees of freedom, which indeed has been found in \cite{Blasi:2022mbl}: the theory with the scalar gauge condition \eqref{scalargaugecond} is not defined in the limit of pure LG. The alternative, which we consider in this paper, is to
focus on the left hand side of \eqref{sym}, where the same symmetric rank-2 tensor field of LG appears, and decide to adopt the same vectorial gauge condition \eqref{vectorgaugecond} as LG. In doing so, two questions should be answered: 
\begin{enumerate}
\item is the vector gauge condition a good gauge fixing, or, equivalently, do the propagators exist in the pure fractonic limit, possibly without being forced to choose a particular gauge~? 
\item in gauge field theory the gauge fixing condition serves to eliminate the redundant degrees of freedom which render infinite the Green functions' generating functional $Z[J]$
\be
Z[J]=\int {\cal D}h_{\mu\nu}\ e^{iS_{inv}+\int J^{\mu\nu}h_{\mu\nu}}\ .
\label{Z}\ee
Do the fact of imposing four (vector) gauge conditions instead of one (scalar) affect the number of physical degrees of freedom of the whole theory $S_{inv}$ ? 
\end{enumerate}
The above are legitimate and well posed questions and, naively, one might answer positively to both. To the first simply because four conditions are more than one, and one expects that they are more than enough to invert the gauge fixed action to find the propagators; to the second for the same reason: four conditions are more than one, and hence the degrees of freedom which are eliminated are too much and differ from those ``killed'' by the scalar gauge choice \eqref{scalargaugecond}. We shall see that the propagators are not singular in the limits of pure fractons or pure LG and that the number of physical degrees of freedom is the same for the two gauge fixing choices \eqref{scalargaugecond} and \eqref{vectorgaugecond}, which therefore are equivalent, with the advantage that the vectorial choice \eqref{vectorgaugecond} does not constrain us to the Landau gauge and allows to easily recover LG.
The paper is organized as follows. In Section \ref{sec:the-model}, starting from the symmetry \eqref{sym}, the action of the theory is derived, which consists of two terms: LG and a fractonic term. The vector gauge condition \eqref{vectorgaugecond} is realized by adding a gauge fixing term to the action. In Section \ref{sec:propagators} the propagators are computed, and the singularities are studied, which correspond to particular phases of the theory. In Section \ref{sec:degrees-of-freedom} we study the degrees of freedom, and we verify that their counting coincides with the known one \cite{Blasi:2022mbl}, without the drawback of being confined to the Landau gauge, which reassures of the fact that the number of degrees of freedom does not depend on a particular choice and that the alternative vectorial gauge fixing condition \eqref{vectorgaugecond} is, indeed, a good one. In Section \ref{sec:summary-and-discussion} we discuss our results.

\section{The model}\label{sec:the-model}

Let us consider the four-dimensional (4D) theory of a symmetric tensor field $h_{\mu\nu}(x)$ which transforms as \eqref{sym}.
The choice \eqref{difftofract} is motivated by the fact that the theory defined by the symmetry \eqref{sym} describes the so called ``fractons''. 
The most general action invariant under \eqref{sym} is
\be
S_{inv}(g_1,g_2)=g_1 S_{LG} + g_2 S_{fract}\ ,
\label{Sinv}\ee
where
\bea
S_{LG} &=&
\int d^4x \left(
	- h \partial^2 h + h_{\mu\nu} \partial^2 h^{\mu\nu} 
	+2 h\partial_\mu \partial_\nu h^{\mu\nu}
	-2 h^{\mu\nu}\partial^\rho \partial_\mu h_{\nu\rho}
	\right)
\label{SLG}\\
S_{fract} &=&
\int d^4x \left(
h^{\mu\nu}\partial^\rho \partial_\mu h_{\rho\nu} - h_{\mu\nu} \partial^2 h^{\mu\nu}\right)
\label{Sfract}\ ,
\eea
and $h(x)$ is the Minkowskian trace of $h_{\mu\nu}(x)$
\be
h=\eta^{\mu\nu}h_{\mu\nu}\ .
\label{A}\ee
The action \eqref{Sinv} appears to be the linear combination of two terms,  which we recognize to be the action $S_{LG}$ of Linearized Gravity (LG) \cite{Carroll.book,Hinterbichler:2011tt} and the pure covariant fractonic action described in \cite{Blasi:2022mbl,Bertolini:2022ijb}. The actions $S_{LG}$ and $S_{fract}$ are separately invariant under \eqref{sym}
\be
\delta S_{LG} = \delta S_{fract} = 0\ ,
\label{}\ee
and $g_{1,2}$ are constants, on which we will come back in a moment. Notice that while the space of 4D local integrated functionals invariant under \eqref{sym} is the linear combination \eqref{Sinv} of two elements, the infinitesimal diffeomorphism symmetry \eqref{diff} uniquely determines one functional only: the LG action $S_{LG}$ \eqref{SLG}
\be
\delta_{diff}S_{LG} = 0\ ,
\label{}\ee
under which the fractonic action $S_{fract}$ \eqref{Sfract} is not invariant
\be
\delta_{diff}S_{fract}= 
2\int d^4x\  \partial^\mu\xi^\nu \left(\partial^\rho\partial_\mu h_{\nu\rho} + \partial^\rho\partial_\nu h_{\rho\mu} - 2 \partial^2 h_{\mu\nu}\right)
\neq 0\ .
\label{}\ee
In other words, the ``fractonic'' symmetry \eqref{sym} is less constraining than the diffeomorphism transformation \eqref{diff}, of which it is a particular case. The action \eqref{Sinv} actually depends on one constant only, because of the possibility of redefining the gauge field by a multiplicative constant $C$  without affecting the physical content of the theory: $h_{\mu\nu}\rightarrow Ch_{\mu\nu}$. Nevertheless, 
we will keep both $g_1$ and $g_2$, in order to track the contributions of the gravitational ($g_2\rightarrow 0$) and the fractonic ($g_1\rightarrow 0$) parts in the rest of the paper. The fact that a quadratic, Lorentz and gauge invariant theory depends on one unavoidable constant is quite uncommon, if not unique. It is not even clear how to call this constant, since it  cannot be a ``coupling'' constant, being the theory $S_{inv}$ free and non interacting, nor a mass, being dimensionless. This peculiarity originates from the fact that the space of functionals invariant under the transformation \eqref{sym} has dimension two, instead of one as it commonly happens. To our knowledge, the only exception is given by the 3D Maxwell-Chern-Simons theory \cite{Deser:1981wh}, which depends on one ``true'' constant as well, but in that case the constant can be identified as a mass. We might say that gravitons may exist alone, while fractons necessarily come with gravitons, lacking, up to now, a symmetry which uniquely determines them. This claim can be found in the fracton Literature \cite{Xu:2006,Gu:2009jh,Xu:2010eg,Pretko:2017fbf}, but it is immediately apparent from the field theoretical point of view.
 Concerning the gauge fixing, the standard way to realize the condition \eqref{vectorgaugecond} is to add to the invariant action \eqref{Sinv} the gauge fixing term
\be
S_{gf}(\xi,\kappa)= 
-\frac{1}{2\xi}\int d^4x \left(\partial^\nu h_{\mu\nu} +\kappa\partial_\mu h\right)^2\ ,
\label{SgfFP}\ee
as it has been done in \cite{Blasi:2015lrg,Blasi:2017pkk,Gambuti:2021meo,Gambuti:2020onb} for LG alone. In a fully equivalent way, \eqref{SgfFP} can be linearized by means of 
a Lagrange multiplier $b^\mu(x)$, also known as Nakanishi-Lautrup field \cite{Nakanishi:1966zz,Lautrup:1967zz}~:
\begin{equation}
S_{gf}(\xi,\kappa)= 
\int d^4x \left[ b^\mu\left(\partial^\nu h_{\mu\nu} +\kappa\partial_\mu h\right)+ \frac{\xi}{2}b_\mu b^\mu\right]
\ .
\label{Sgf}\end{equation}
In $S_{gf}(\xi,\kappa)$ two gauge fixing parameters appear: $\xi$ and $\kappa$. The first -$\xi$- governs the type of gauge fixing. For instance $\xi=0$ and $\xi=1$ are respectively the Landau and Feynman gauges. The second -$\kappa$- tunes the type of gauge fixing picked up by $\xi$. For instance, the Landau gauge in LG corresponds to a class of gauge choices, and it is realized by $\xi=0$ and generic $\kappa$ ($\kappa=\frac{1}{2}$ being the harmonic Landau gauge \cite{Carroll.book,Gambuti:2020onb}).

\section{Propagators}\label{sec:propagators}

To find the propagators of the theory, we write the gauge fixed action 
\be
S(g_1,g_2;\xi,\kappa) = S_{inv}(g_1,g_2) + S_{gf}(\xi,\kappa)
\label{gaugefixedaction}\ee
in momentum space~:
\be
\int d^4p
		\begin{pmatrix}
			\tilde{h}^{\mu\nu}(p) &
			\tilde{b}^\gamma(p)
		\end{pmatrix}
		\begin{pmatrix}
			\tilde{\Omega}_{\mu\nu,\alpha\beta}(p) & \tilde{\Lambda}^*_{\mu\nu,\lambda}(p)\\
			\tilde{\Lambda}_{\gamma,\alpha\beta}(p) & \tilde{H}_{\gamma\lambda}(p)
		\end{pmatrix}
		\begin{pmatrix}
			\tilde{h}^{\alpha\beta}(-p)\\
			\tilde{b}^\lambda(-p)
		\end{pmatrix},
\label{momentumaction}\ee
where $\tilde{\Omega}_{\mu\nu,\alpha\beta}(p)$, $\tilde{\Lambda}^*_{\mu\nu,\lambda}(p)$ and $\tilde{H}_{\gamma\lambda}(p)$ are $p$-dependent tensor operators. The propagators of the theory are obtained by inverting the operator matrix appearing in \eqref{momentumaction}. In order to do this, it is useful to write $\tilde{\Omega}_{\mu\nu,\alpha\beta}(p)$, $\tilde{\Lambda}_{\alpha\beta,\mu}(p)$ and $\tilde{H}_{\mu\alpha}(p)$ on the corresponding tensorial basis, as follows
\bea
\tilde{\Omega}_{\mu\nu,\alpha\beta} &= &
\tilde{t}A_{\mu\nu,\alpha\beta} + \tilde{u}B_{\mu\nu,\alpha\beta}+ 
\tilde{v}C_{\mu\nu,\alpha\beta}+ \tilde{z}D_{\mu\nu,\alpha\beta} + 
\tilde{w}E_{\mu\nu,\alpha\beta}
\label{Omega}\\
\tilde{\Lambda}_{\alpha\beta,\mu} &=&
 -\frac{i}{2}\left[\tilde{f}(d_{\alpha\mu} p_\beta + d_{\beta\mu}p_\alpha) + \tilde{g}d_{\alpha\beta} p_\mu + \tilde{l}e_{\alpha\beta}p_\mu \right]
 \label{Lambda}\\
\tilde{H}_{\mu\alpha} &=& \tilde{r} d_{\mu\alpha} + \tilde{s} e_{\mu\alpha}\ ,
\label{H}\eea
where $e_{\mu\nu}(p)$ and $d_{\mu \nu}(p)$ are transverse and longitudinal projectors, respectively,
\be
e_{\mu\nu}=\frac{p_\mu p_\nu}{p^2}\ \ ;\ \ 
d_{\mu\nu}=\eta_{\mu\nu}-e_{\mu\nu}\ ,
\label{projectors}\ee
and the rank-4 tensor $\tilde{\Omega}_{\mu\nu,\alpha\beta}(p)$ \eqref{Omega} is expanded on a basis of operators 
\be
X_{\mu\nu,\alpha\beta}\equiv (A,B,C,D,E)_{\mu\nu,\alpha\beta}
\label{baseX}\ee
which can be found in Appendix \ref{sec:basis-for-the-omega-tensors}, together with their properties. The coefficients appearing in \eqref{Omega}, \eqref{Lambda} and \eqref{H} are found to be
	\begin{align}
		\tilde{t} &= (2g_1+g_2)p^2			&	\tilde{u} &= 0	&	\tilde{v} &= (g_2-g_1) p^2	&	\tilde{z} &= \frac{1}{2}g_2 p^2	&	\tilde{w} &= 0\label{3.8}\\[5pt]
		\tilde{f} &= \frac{1}{2}		&	\tilde{g} &= \kappa	&	\tilde{l} &= 1+\kappa	&	\tilde{r} &= \frac{\xi}{2}		&	\tilde{s} &= \frac{\xi}{2}\label{3.9}\ .
	\end{align}
The propagators of the theory are organized in a matrix of tensor operators as well
\begin{equation}\label{matricepropagatori}
		\begin{pmatrix}
			\hat{{G}}^{\alpha\beta,\rho\sigma}(p) & \hat{{G}}^{\alpha\beta,\tau}(p)\\
			\hat{{G}}^{*\lambda,\rho\sigma}(p) & \hat{{G}}^{\lambda\tau}(p)
		\end{pmatrix}\ ,
	\end{equation}
	where
\begin{equation}
		\hat{G}_{\alpha\beta,\rho\sigma}(p) = \left\langle \tilde{h}_{\alpha\beta}(p)\tilde{h}_{\rho\sigma}(-p)\right\rangle =  \hat{t}A_{\alpha\beta,\rho\sigma} + 
		\hat{u}B_{\alpha\beta,\rho\sigma} + \hat{v}C_{\alpha\beta,\rho\sigma} + \hat{z}D_{\alpha\beta,\rho\sigma} + \hat{w}E_{\alpha\beta,\rho\sigma}
\label{hhprop}	\end{equation}
	\begin{equation}\label{hbprop}
		\hat{G}_{\alpha\beta,\rho}(p) = \left\langle \tilde{h}_{\alpha\beta}(p)\tilde{b}_{\rho}(-p)\right\rangle =  i\left[\hat{f}(d_{\alpha\rho} p_\beta + d_{\beta\rho}p_\alpha) + \hat{g}d_{\alpha\beta} p_\rho +\hat{l}e_{\alpha\beta}p_\rho\right]
	\end{equation}
	\begin{equation}\label{bbprop}
		\hat{G}_{\alpha\rho}(p) = \left\langle \tilde{b}_{\alpha}(p)\tilde{b}_{\rho}(-p)\right\rangle = \hat{r}d_{\alpha\rho} + \hat{s}e_{\alpha\rho}\ ,
	\end{equation}
and the set of coefficients 
\be
\left\{\hat t\,,\, \hat u\,,\, \hat v\,,\, \hat z\,,\, \hat w\,,\, \hat f\,,\, \hat g\,,\, \hat l\,,\, \hat r\,,\, \hat s\right\}
\label{propcoeff}\ee
are determined by the request that the matrix of propagators \eqref{matricepropagatori} satisfies
\begin{equation}\label{eq:matriciale}
		\begin{pmatrix}
			\tilde{\Omega}_{\mu\nu,\alpha\beta} & \tilde{\Lambda}^*_{\mu\nu,\lambda}\\
			\tilde{\Lambda}_{\gamma,\alpha\beta} & \tilde{H}_{\gamma\lambda}
		\end{pmatrix}
		\begin{pmatrix}
			\hat{G}^{\alpha\beta,\rho\sigma} & \hat{G}^{\alpha\beta,\tau}\\
			\hat{G}^{*\lambda,\rho\sigma}& \hat{G}^{\lambda\tau}
		\end{pmatrix}
		=
		\begin{pmatrix}
			\mathcal{I}_{\mu\nu}^{\ \ \rho\sigma} &0 \\
			0 & \delta_{\gamma}^{\ \tau}
		\end{pmatrix}\ ,
	\end{equation}
where $\mathcal{I}_{\mu \nu, \alpha\beta}$ is the rank-4 tensor identity 
\be
\mathcal{I}_{\mu \nu, \rho \sigma} = \frac{1}{2} (\eta_{\mu \rho} \eta_{\nu \sigma} + \eta_{\mu \sigma} \eta_{\nu \rho})\ .
\label{identity}\ee
In Appendix \ref{sec:2g1g2neq0} we show that the four tensor equations \eqref{eq:matriciale} are solved by
	\begin{align}
		&\hat{t} = \frac{(4\kappa+1)}{(\kappa+1)(2g_1+g_2)p^2}\label{eq:that}\\*
		&\hat{u} = \frac{\kappa(4\kappa+1)-2\xi(2g_1+g_2)}{(\kappa+1)^2(2g_1+g_2)p^2}\label{eq:uhat}\\*
		&\hat{v} = \frac{1}{(g_2-g_1)p^2}\label{eq:vhat}\\*
		&\hat{z} = \frac{4\xi}{(2\xi g_2 -1)p^2}\label{eq:zhat}\\*
		&\hat{w} = \frac{-4\kappa}{(\kappa+1)(2g_1+g_2)p^2}\label{eq:what}\\*
		&\hat{f} = \frac{-2}{(2\xi g_2-1)p^2}\label{eq:fhat}\\
		&\hat{g} = 0\label{eq:ghat}\\
		&\hat{l} = \frac{2}{(\kappa+1)p^2}\label{eq:lhat}\\
		&\hat{r} = \frac{4g_2}{(2\xi g_2-1)}\label{eq:rhat}\\
		&\hat{s} = 0\label{eq:shat}\ .
	\end{align}
Now we can answer the first of our questions, concerning the fractonic limit: the vector gauge fixing condition \eqref{vectorgaugecond} is a good one not only for LG ($g_2=0$), but also for the pure fractonic case ($g_1=0$), as expected. On the other hand, we see that the propagators are singular in four cases~:
\bea
2g_1+g_2 &=& 0 \label{2g1+g2}\\
g_1-g_2 &=& 0\label{g1-g2}\\
2\xi g_2-1 &=& 0 \label{2kappag2-g1}\\
\kappa+1 &=& 0 \label{kappa+1}
\ .
\eea
For what concerns the singularity \eqref{kappa+1}, it simply implies that the ``secondary'' gauge fixing parameter $\kappa$, which tunes the gauge fixing choice of the ``primary'' parameter $\xi$ in $S_{gf}$ \eqref{Sgf}, should be 
\be
\kappa\neq -1\ ,
\label{kneq-1}\ee
as it happens also in LG \cite{Blasi:2015lrg,Gambuti:2020onb,Blasi:2017pkk,Gambuti:2021meo}. The remaining singularities involve the action parameters $g_1$ and $g_2$ and the gauge fixing parameter $\xi$. The poles in the propagators give us information on the structure of the theory. For instance, they might signal the presence of masses, possibly not standard, as in the topologically massive Maxwell-Chern-Simons theory \cite{Deser:1981wh}. Or they might indicate the presence of phase transitions in the theory, like in QCD \cite{Cabibbo:1975ig,Halasz:1998qr}, or in the sigma model \cite{Baym:1977qb,Grater:1994qx}, or in the fracton theory itself \cite{Blasi:2022mbl}.  Therefore, the singularities appearing in the propagators of the theory $S_{inv}$ should be treated separately and with care.
 \begin{itemize}
 \item${\bf g_1=g_2}$\hfill \\
In this case, after a field redefinition, the invariant action \eqref{Sinv}  is 
	\begin{equation}\label{g1=g2}
\left.S_{inv}(g_1,g_2)\right|_{g_1=g_2}
= \int d^4p\ (\tilde{h} p^2 \tilde{h} -2 \tilde{h} p_\mu p_\nu \tilde{h}^{\mu\nu} +\tilde{h}^{\mu\nu}p^\rho p_\mu \tilde{h}_{\nu\rho})\ .
	\end{equation}
Notice that defining	
	\begin{equation}
		\tilde{H}^\rho \equiv p^\rho\tilde{h} - p_\nu \tilde{h}^{\rho\nu}\ ,
	\end{equation}
the action \eqref{g1=g2} trivializes into	
	\begin{equation}
	\left.S_{inv}(g_1,g_2)\right|_{g_1=g_2} = - \int d^4p\ \tilde{H}^\rho \tilde{H}_\rho\ ,
	\end{equation}
which does not contain any kinetic term. Hence, the singularity at $g_1=g_2$ is explained as a point where the theory trivializes and does not propagate, and this case will be excluded from now on.
\item${\bf 2g_1+g_2=0}$\hfill \\
In this case the invariant action \eqref{Sinv}, after a field redefinition,  reads 
	\begin{equation}\label{2g1+g_2}
\left.S_{inv}(g_1,g_2)\right|_{2g_1+g_2=0}
= \int d^4p\ 
 \left(\tilde{h} p^2 \tilde{h} -3\tilde{h}_{\mu\nu} p^2 \tilde{h}^{\mu\nu} -2 \tilde{h} p_\mu p_\nu \tilde{h}^{\mu\nu} +4 \tilde{h}^{\mu\nu}p^\rho p_\mu \tilde{h}_{\nu\rho} \right)\ .
	\end{equation}
We see that with this choice the action does not depend on the trace $\tilde h (p)$. In fact, defining
\begin{equation}
		{\bar{h}}_{\mu\nu}(p)\equiv \tilde{h}_{\mu\nu}(p)-\frac{1}{4}\eta_{\mu\nu}\tilde{h}(p)\ ,
\label{tracelessh}	\end{equation}
with
	\begin{equation}
		{\bar{h}}(p) = 0\ ,
	\end{equation}
the action \eqref{2g1+g_2} can be written in terms of $\bar{h}_{\mu\nu}(p)$ only~:
\begin{equation}\label{eq:tracelessaction}
		\left.S_{inv}(g_1,g_2)\right|_{2g_1+g_2=0} = \int d^4p\ \left(-3{\bar{h}}_{\mu\nu}p^2{\bar{h}^{\mu\nu}}+4{\bar{h}^{\mu\nu}}p^\rho p_\mu {\bar{h}_{\nu\rho}}\right)\ .
	\end{equation}
Hence, in this case the theory is traceless, and the singularity at the point $2g_1+g_2=0$ indicates a change in the counting of the degrees of freedom, as we shall explicitly show. The gauge fixing term \eqref{Sgf} does not depend on the trace $h(x)$ anymore and, hence, on the gauge fixing parameter $\kappa$. In momentum space it reads
	\begin{equation}\label{Sgftraceless}
S_{gf}(\xi)=
 \int d^4p\ \left(-i\tilde{b}_\mu p_\nu \bar{h}^{\mu\nu} + \frac{\xi}{2}\tilde{b}_\mu \tilde{b}^\mu\right)\ .
	\end{equation}
The propagators of the traceless theory are well defined, and the coefficients, computed in Appendix \ref{sec:2g1g20}, are 	
	\begin{align}
		\hat{t} &= - \frac{1}{3p^2}	&\hat{u} &= \frac{2\xi}{(2\xi-1)p^2}\\
		\hat{v}&= - \frac{1}{3p^2}	&\hat{z} &= \frac{-4\xi}{(4\xi+1)p^2}\\
		\hat{w} &= 0					&\hat{f} &= \frac{2}{(4\xi+1)p^2}\\
		\hat{g} &= 0					& \hat{l} &= \frac{-2}{(2\xi-1)p^2}\\
		\hat{r} &= \frac{8}{(4\xi+1)}	&	\hat{s} &= \frac{4}{(2\xi-1)}\ .
	\end{align}
From the above coefficients we see that the particular values of the primary gauge parameter $\xi=-1/4$ and $\xi=1/2$ should be excluded.
\item${\bf 2\xi g_2-1=0}$\hfill \\	
We first notice that this singularity is not present  in the pure LG case $g_2=0$, as it is readily seen from the coefficients of the propagators \eqref{eq:zhat}, \eqref{eq:fhat} and \eqref{eq:rhat}.
Then, we remark that $g_2$ is a physical parameter, hence cannot depend on $\xi$, which is a gauge, unphysical, parameter. This means that $2\xi g_2-1=0$ should be interpreted as a singularity in $\xi$ as a function of $g_2$, and not $viceversa$. In other words, we shall not exclude values of $g_2$ in order to admit a particular gauge, but, rather, the singularity must be read as a condition on the gauge fixing parameter $\xi$~:
\be 
\xi\neq \frac{1}{2g_2}\ .
\label{kappacond}\ee 
We consider the gauge fixing term \eqref{SgfFP}, before the the introduction of the Lagrange multiplier $b^\mu(x)$. At $2\xi g_2-1=0$ the gauge fixed action is
	\begin{multline}
\left.S(g_1,g_2;\xi,\kappa)\right|_{2\xi g_2-1=0} =
\int d^4p\ 
\left[
		(g_1-g_2\kappa^2)\tilde{h}p^2\tilde{h}
		-(g_1-g_2)\tilde{h}_{\mu\nu}p^2\tilde{h}^{\mu\nu}+\right.\\*\left.
		-2(g_1+g_2\kappa)\tilde{h}p_\mu p_\nu \tilde{h^{\mu\nu}}
		+2(g_1-g_2)\tilde{h}^{\mu\nu}p_\mu p^\rho \tilde{h}_{\nu\rho}\right]\ .
	\end{multline}
Using the general results of Appendix \ref{sec:2g1g2neq0} it is easy to verify that this theory does not have propagators. As a remark, if we choose also $\kappa+1=0$, we find a curious result
	\begin{equation}
\left.S(g_1,g_2;\xi,\kappa)\right|_{2\xi g_2-1=0,\kappa+1=0}
=(g_1-g_2)S_{LG}\ .
\label{nogf}	\end{equation}
This means that with the particular gauge choice which involves both the singularities in the two gauge parameters $\xi$ and $\kappa$,  the gauge fixing procedure fails in choosing one representative for each gauge orbit, which is what the gauge fixing is supposed to do. In fact, according to \eqref{nogf}, in this particular gauge, the fracton contribution disappears, and the gauge fixed action coincides with $S_{LG}$ alone, which still needs to be gauge fixed. It also appears the fact which we already know that for $g_1=g_2$, which is the trivial, non propagating case already considered, the action vanishes. We thus showed that for $\xi=1/2g_2$ and $\kappa=-1$ the gauge fixed action $S(g_1,g_2;\xi,\kappa)$ coincides with the invariant, not gauge fixed, action $S_{LG}$.

\end{itemize}

\section{Degrees of freedom}\label{sec:degrees-of-freedom}

The counting of the degrees of freedom of the theory described by the action $S_{inv}$ \eqref{Sinv} is a crucial point. This is true in general, but for this paper it is even more true. What we already know from \cite{Blasi:2022mbl} is that, if we adopt the standard scalar gauge choice \eqref{scalargaugecond}, which is the natural one when dealing with a gauge transformation depending on a scalar parameter, the degrees of freedom turn out to be six, as in LG alone (five in the traceless case), as if the fractonic contribution $S_{fract}$ \eqref{Sfract} was not present. However, in \cite{Blasi:2022mbl} the choice of the Landau gauge appears to be mandatory, which is rather unpleasant, although the physical results should not depend on the gauge choice. Still, it would be preferable to avoid such a restriction, which instead seems unavoidable in the scalar gauge. This, together with the fact that LG cannot be reached as a limit, leads us to conclude that the scalar gauge choice is not that natural, as the direct application of the Faddeev-Popov procedure suggests. The aim of this paper is to see whether the alternative and, at first sight, exotic choice of the $vector$ gauge condition \eqref{vectorgaugecond} is an acceptable, and possibly better, one. In the previous Section we passed the first test: we have seen that the vector gauge condition leads to well defined propagators, with a pole which corresponds to the traceless theory, in accordance to the scalar case \cite{Blasi:2022mbl}. In this Section we face the trickier point of the counting of the degrees of freedom. Not only we should recover the known result, but, and more important, we should show that the number of degrees of freedom does not depend on the gauge choice, which was impossible with the scalar gauge condition \eqref{scalargaugecond} frozen in the Landau gauge. This fact is not obvious, since the role of the gauge fixing is to eliminate the redundant degrees of freedom, in order to render finite the path integral $Z[J]$  \eqref{Z}, and the justified fear is that the four conditions represented by the vector choice \eqref{vectorgaugecond} might lead to underestimate the degrees of freedom with respect to the unique scalar condition \eqref{scalargaugecond}.
The usual way to count the degrees of freedom is to look for the constraints deriving from the equations of motion of the gauge fixed action
\be
S(g_1,g_2;\xi,\kappa)= 
S_{inv}(g_1,g_2) + S_{gf}(\xi,\kappa)\ ,
\label{Stot}\ee
where the invariant action $S_{inv}(g_1,g_2)$ and the gauge fixing term $S_{gf}(\xi,\kappa)$ are given by \eqref{Sinv} and \eqref{Sgf}, respectively. In momentum space, we get
\bea
\frac{\delta S}{\delta \tilde{h}^{\mu\nu}} 
&=& 
2g_1\eta_{\mu\nu}p^2\tilde{h} + 2(g_2-g_1)p^2\tilde{h}_{\mu\nu}-2g_1\eta_{\mu\nu}p_\alpha p_{\beta}\tilde{h}^{\alpha\beta}-2g_1p_\mu p_\nu\tilde{h} +(2g_1-g_2)p^\alpha(p_\mu  \tilde{h}_{\alpha\nu} +p_\nu  \tilde{h}_{\alpha\mu}) 
\nonumber\\
&& +\frac{i}{2}(p_\nu\tilde{b}_\mu+p_\mu\tilde{b}_\nu)+i\kappa\eta_{\mu\nu}p_\alpha \tilde{b}^{\alpha} = 0 \label{eomh}
\\
\frac{\delta S}{\delta \tilde{b}^{\mu}} &=& -ip^\alpha \tilde{h}_{\alpha\mu} -i\kappa p_\mu\tilde{h} + \xi\tilde{b}_\mu = 0\ . \label{eomb}
\eea
If our task was just to count the degrees of freedom, given that they must not depend on the gauge choice, we would immediately find the result by choosing $\xi=\kappa=0$ in $S_{gf}$ \eqref{Sgf}, which belongs to the class of Landau gauges. The $\tilde b^\mu$-equation of motion \eqref{eomb} gives
\be
p^\alpha \tilde{h}_{\alpha\mu}=0\ ,
\label{xi=k=0}\ee
which are the four constraints needed to recover the six degrees of freedom (five in the traceless case) which we expect for the symmetric rank-2 tensor field $h_{\mu\nu}(x)$. But we want more, that is to show that this number does not depend on the gauge choice, which would render the vector gauge condition a good one under any respect. Achieved that, the vector choice would be even preferable to the scalar one, since the Landau gauge would not be the only possibility, and LG could be obtained as a limit. Hence we proceed without choosing a particular gauge, and
we saturate \eqref{eomh} with $\eta^{\mu\nu}$, $e^{\mu\nu}(p)$ \eqref{projectors} and $p^\mu$~:
\begin{flalign}{}
		&\eta^{\mu\nu}\frac{\delta S}{\delta \tilde{h}^{\mu\nu}} = 2(2g_1+g_2)\left(p^2\tilde{h}-p_\alpha p_{\beta}\tilde{h}^{\alpha\beta}\right)+i(1+4\kappa)p_\alpha \tilde{b}^{\alpha} = 0\label{eq:satura1}\\
		&e^{\mu\nu}\frac{\delta S}{\delta \tilde{h}^{\mu\nu}} = i(1+\kappa)p_\alpha \tilde{b}^{\alpha} = 0\label{eq:satura2}\\
		&p^\nu\frac{\delta S}{\delta \tilde{h}^{\mu\nu}} = 2g_2p^\alpha \left(p^2 \tilde{h}_{\alpha\mu} -p_\mu  p^\beta\tilde{h}_{\alpha\beta}\right) + ip^2\tilde{b}_\mu +i(1+2\kappa)p_\mu p_\alpha \tilde{b}^{\alpha} = 0\ .\label{eq:satura3}
	\end{flalign}
Multiplying \eqref{eomb} by $p^\mu$, we get
	\begin{equation}
		p^\mu \frac{\delta S}{\delta \tilde{b}^{\mu}} = ip_\alpha p_\beta\tilde{h}^{\alpha\beta} +i\kappa p^2\tilde{h} - \xi p_\alpha \tilde{b}^{\alpha} = 0\ .\label{eq:satura4}
	\end{equation}
We separately study  the generic case $2g_1+g_2\neq0$ and the traceless case $2g_1+g_2=0$\ .

\subsection{Case $2g_1+g_2\neq0$}\label{sec:case-2g1g2neq0}

From \eqref{eq:satura2}, remembering that $\kappa+1\neq0$, we get the condition
	\begin{equation}
		p_\alpha \tilde{b}^\alpha =0\ ,
	\end{equation}
which, plugged in \eqref{eomb}, \eqref{eq:satura1}, \eqref{eq:satura3} and \eqref{eq:satura4}, yields
	\begin{align}
		&\xi\tilde{b}_\mu = i\left(p^\alpha \tilde{h}_{\alpha\mu} +\kappa p_\mu\tilde{h} \right)\label{eq:Bl}\\
		&(2g_1+g_2)\left(p^2\tilde{h}-p_\alpha p_{\beta}\tilde{h}^{\alpha\beta}\right) = 0\label{eq:satura12}\\
		&2g_2p^\alpha \left(p^2 \tilde{h}_{\alpha\mu} -p_\mu p^\beta\tilde{h}_{\alpha\beta}\right) + ip^2\tilde{b}_\mu = 0\label{eq:satura32}\\
		&ip_\alpha p_\beta\tilde{h}^{\alpha\beta} +i\kappa p^2\tilde{h} = 0\ .\label{eq:satura42}
	\end{align}
Now, since we are outside the critical point $2g_1+g_2=0$ \eqref{2g1+g2}, from \eqref{eq:satura12} and \eqref{eq:satura42} we have
\bea
		p^2\tilde{h}&=&p_\alpha p_{\beta}\tilde{h}^{\alpha\beta}\label{eq:p2hpph}\\
		-\kappa p^2\tilde{h} &=& p_\alpha p_\beta\tilde{h}^{\alpha\beta}\ ,
\eea
that is
\begin{equation}
		(1+\kappa)p^2\tilde{h} =0\Rightarrow p^2\tilde{h} =0\ ,
\label{p2h=0}	\end{equation}
hence
\begin{equation}\label{eq:pcurrent}
		p_\alpha p_{\beta}\tilde{h}^{\alpha\beta} = 0\ .
\end{equation}
Notice that the conditions \eqref{p2h=0} and \eqref{eq:pcurrent} are the ones holding for LG alone ($g_2=0$) \cite{Blasi:2017pkk,Gambuti:2021meo}. It appears, therefore, that the fracton contribution ($g_1=0$) \eqref{Sfract} to the total invariant action \eqref{Sinv} is irrelevant as far as the degrees of freedom are concerned, which is an unexpected result. Nonetheless we can directly check  this result. Substituting the conditions \eqref{p2h=0} and \eqref{eq:pcurrent} into \eqref{eq:satura32}, we find
	\begin{equation}\label{eq:misfatto}
		-ip^2\tilde{b}_\mu = 2g_2 p^2p^\alpha \tilde{h}_{\alpha\mu}\ .
	\end{equation}
Using \eqref{eq:Bl} and \eqref{p2h=0}, assuming $\xi\neq0$, $i.e.$ excluding for the moment the Landau gauge, equation \eqref{eq:misfatto} becomes
	\begin{equation}
		(2g_2\xi-1)p^2p^\alpha\tilde{h}_{\alpha\mu} =0\ .
	\end{equation}
We previously studied the case $2g_2\xi-1=0$, which we now exclude. This means that
	\begin{equation}\label{eq:ppph}
		p^2p^\alpha\tilde{h}_{\alpha\mu} = 0
	\end{equation}
	and, from \eqref{eq:misfatto},
	\begin{equation}
		p^2\tilde{b}_\mu = 0\ .
	\end{equation}
We now define
	\begin{equation}
		\tilde{J}_\alpha \equiv p^\beta\tilde{h}_{\alpha\beta}\ ,
\label{defJ}	\end{equation}
which, because of \eqref{eq:pcurrent}, is a conserved current
	\begin{equation}
		p^\alpha\tilde{J}_\alpha = 0\ .
\label{pJ}	\end{equation}
The solution of \eqref{pJ} is
\begin{equation}\label{eq:jB}
		\tilde{J}_\alpha = \epsilon_{\alpha\mu\nu\rho}p^\mu \tilde{B}^{\nu\rho}\ ,
	\end{equation}
where $\tilde{B}^{\nu\rho}$ is a generic antisymmetric tensor. Plugging \eqref{defJ} in \eqref{eq:ppph}, we have
\begin{equation}
p^2\tilde{J}_\alpha = p^2\epsilon_{\alpha\mu\nu\rho}p^\mu \tilde{B}^{\nu\rho} = 0\ ,
	\end{equation}
	and therefore
	\begin{equation}
		p^2\tilde{B}_{\rho\lambda} = p_\rho\tilde{B}_\lambda - p_\lambda \tilde{B}_\rho\ .
	\end{equation}
From \eqref{eq:jB} we then deduce that the current $J_\alpha$ vanishes 		\begin{equation}\label{eq:lastconstr}
		\tilde{J}_\alpha = p^\beta\tilde{h}_{\alpha\beta} = 0\ .
	\end{equation}
We now come back to the Landau gauge $\xi=0$, which has been excluded in achieving the above result. We now show that \eqref{eq:lastconstr} holds also in this case. The $\tilde{b}^\mu$-equation of motion in the Landau gauge is
	\begin{equation}
		p^\alpha \tilde{h}_{\alpha\mu} = -\kappa p_\mu\tilde{h}\ .
	\end{equation}
The degrees of freedom must not depend on the gauge choice. We therefore choose $\kappa=0$ and we get 
	\begin{equation}\label{eq:llgdof}
		{p^\alpha \tilde{h}_{\alpha\mu} = 0}\ ,
	\end{equation}
which represents four constraints on the 4D symmetric tensor field $\tilde h_{\mu\nu}(p)$. Hence the number of degrees of freedom are six, at least if $2g_1+g_2\neq0$. This coincides with the number of degrees of freedom of LG alone \cite{Blasi:2017pkk,Gambuti:2021meo}. 
We have seen that $2g_1+g_2=0$ corresponds to the traceless case, which is interesting and will be treated separately. 

\subsection{Case $2g_1+g_2=0$}\label{sec:case-2g1g20}

As we have seen in Section \ref{sec:propagators}, this case corresponds to the traceless theory. The gauge fixed action can be written in terms of the traceless field $\bar h_{\mu\nu}(x)$ \eqref{tracelessh}, with $\kappa=0$, and, in momentum space, it reads 
	\begin{equation}
	\left.S(g_1,g_2;\xi,\kappa)\right|_{2g_1+g_2=0,\kappa=0}= \int d^4p\ \left(
		-3\bar{h}_{\mu\nu}p^2\bar{h}^{\mu\nu}
		+4\bar{h}^{\mu\nu}p_\mu p^\rho \bar{h}_{\nu\rho} 
		-i\tilde{b}_\mu p_\nu \bar{h}^{\mu\nu}
		+\frac{\xi}{2}\tilde{b}_\mu\tilde{b}^\mu
		\right)\ ,
	\end{equation}
whose equations of motion are
\bea
		\frac{\delta S}{\delta \bar{h}^{\mu\nu}}  &=& 
		-6p^2\bar{h}_{\mu\nu} 
		+4p^\alpha\left(p_\mu\bar{h}_{\alpha\nu}+p_\nu \bar{h}_{\alpha\mu}\right)
		+\frac{i}{2}\left(p_\nu\tilde{b}_\mu +p_\mu\tilde{b}_\nu\right) = 0
	\label{eq:1}\\
		\frac{\delta S}{\delta \tilde{b}^\mu} &=&  -ip^\nu \bar{h}_{\mu\nu}+\xi\tilde{b}_\mu =0\ .
\label{eq:2}\eea
Saturating \eqref{eq:1} with $\eta^{\mu\nu}$, $e^{\mu\nu}(p)$ and $p^\nu$, we get
\bea
\eta^{\mu\nu}\frac{\delta S}{\delta \bar{h}^{\mu\nu}} &=& 8p_\alpha p_\beta \bar{h}^{\alpha\beta} +ip_\alpha \tilde{b}^\alpha =0\label{eq:sat1} \\
e^{\mu\nu}\frac{\delta S}{\delta \bar{h}^{\mu\nu}} &=& 2p_\alpha p_\beta\bar{h}^{\alpha\beta} +ip_\alpha\tilde{b}^\alpha =0\label{eq:sat2}\\
p^\nu\frac{\delta S}{\delta \bar{h}^{\mu\nu}} &=&-2p^2 p^\alpha \bar{h}_{\alpha\mu} + 4p_\mu p_\alpha p_\beta \bar{h}^{\alpha\beta}+\frac{i}{2}p^2\tilde{b}_\mu +\frac{i}{2}p_\mu p_\alpha \tilde{b}^\alpha =0\ .\label{eq:sat3}
\eea
From \eqref{eq:sat1} and \eqref{eq:sat2} we have
\bea 
p_\alpha p_\beta \bar{h}^{\alpha\beta} &=& 0 \label{eq:key2}\\
p_\alpha\tilde{b}^\alpha &=&0\ .
\eea
If $\xi\neq0$, the Lagrange multiplier $\tilde b_\mu(p)$ can be obtained from \eqref{eq:2} and, plugged in \eqref{eq:sat3}, using \eqref{eq:key2}, we get
	\begin{equation}
		(4\xi+1)p^2p^\alpha \bar{h}_{\alpha\mu} = 0\ .
	\end{equation}
We have already excluded the gauge choice $4\xi+1=0$, which is the propagator singularity $2\xi g_2-1=0$ at $2g_1+g_2=0$, hence
	\begin{equation}\label{eq:key1}
		p^2p^\alpha \bar{h}_{\alpha\mu} = 0\ .
	\end{equation}
Now, using the same argument of Section \ref{sec:case-2g1g2neq0}, involving $\tilde J_\alpha(p)$ \eqref{defJ} and $\tilde B^{\nu\rho}(p)$ \eqref{eq:jB}, Eqs. \eqref{eq:key2} and \eqref{eq:key1} imply the four constraints
	\begin{equation}\label{eq:key3}
		p^\alpha \bar{h}_{\alpha\mu} = 0\ .
	\end{equation}
On the other hand, when $\xi=0$, $i.e.$ in the Landau gauge, the equation of motion of the Lagrange multiplier \eqref{eq:2} directly gives the constraint \eqref{eq:key3}. Hence, in all cases we have four constraints on a traceless rank-2 symmetric tensor. Therefore, when $2g_1+g_2=0$, the degrees of freedom are five. \\

The results for the different values of the action constants $g_1$ and $g_2$ and of the gauge fixing parameters $\xi$ and $\kappa$ are summarized in Table \hyperref[table:dofsummary]{1}.

\begin{table}[H]
	\centering
	\resizebox{1\columnwidth}{!}{%
		\bgroup
		\setlength\tabcolsep{15pt}
		\def\arraystretch{1.5}{
			\begin{tabular}{|c|cccc|}
				\hline
				\multirow{2}{*}{$\mathbf{g_1}$, $\mathbf{g_2}$} & \multicolumn{2}{c|}{{\bf vectorial gauge fixing}} & \multicolumn{2}{c|}{{\bf scalar gauge fixing}}\\ \cline{2-5} 
				& \multicolumn{1}{c|}{\textbf{degrees of freedom}} & \multicolumn{1}{c|}{\textbf{forbidden gauges}} & \multicolumn{1}{c|}{\textbf{degrees of freedom}} & \textbf{forbidden gauges} \\ \hline
				$g_1\neq g_2\neq 0$, $2g_1+g_2\neq0$ & \multicolumn{1}{c|}{6} & \multicolumn{1}{c|}{$\xi = \frac{1}{2g_2}$, $\kappa=-1$} & \multicolumn{1}{c|}{6} & $\xi\neq0$ \\ \hline
				$g_2=0$ (LG) & \multicolumn{1}{c|}{6} & \multicolumn{1}{c|}{$\kappa=-1$} & \multicolumn{2}{c|}{not defined}    \\ \hline
				$2g_1+g_2=0$& \multicolumn{1}{c|}{5} & \multicolumn{1}{c|}{$\xi=\left\{\frac{1}{2}, -\frac{1}{4}\right\}$} & \multicolumn{1}{c|}{5} &  $\xi\neq0$  \\ \hline
				$g_1=g_2$& \multicolumn{4}{c|}{trivial}                                                    \\ \hline
			\end{tabular}
		\egroup}
	}
	\label{table:dofsummary}
	\caption{\small{Summary of results and comparison with the scalar case}}
\end{table}

\section{Summary and discussion}\label{sec:summary-and-discussion}

In this paper we considered the theory of a symmetric rank-2 tensor $h_{\mu\nu}(x)$, invariant under the symmetry \eqref{sym},
which is the covariant extension of the fractonic symmetry studied in the Literature. 
The main novelty of our approach, with respect to the existing Literature concerning fractons, is to give a $covariant$ theory of this new type of quasiparticles, and this is not a formal point. In fact the covariant extension \eqref{sym}  of the fractonic symmetry, which usually involves only space and not time derivatives, leads immediately to the action \eqref{Sinv}, which makes evident the relation between Linearized Gravity and fractons. Moreover, as explained in \cite{Bertolini:2022ijb}, the main results concerning fractons, in particular the existence of tensorial electric and magnetic fields, the Gauss constraint, the Maxwell-like Hamiltonian and the dipole response to ``electromagnetic'' fields through a ``Lorentz force'', to cite a few, are indeed consequences of a covariant action for a symmetric rank-2 tensor field $h_{\mu\nu}(x)$, invariant under the covariant extension of the fracton transformation \eqref{sym},
which therefore plays, as usual in quantum field theory, a central role. This, in our opinion, is the main physical motivation for studying the action \eqref{Sinv}.
Without this strong physical motivation, a theory defined by the transformation  \eqref{sym} suffers of several drawbacks from the field theoretical point view. First,  \eqref{sym} is quadratic in the spacetime derivatives. This implies that, in 4D, the gauge parameter $\Phi(x)$ must have negative mass dimensions. In fact, the most general action invariant under \eqref{sym} is given by $S_{inv}$ \eqref{Sinv}, and power counting tells us that $[h_{\mu\nu}]=1$, whence the unusual negative mass assignment for the gauge parameter. Had we dealt with a field theory exercise, it would have been better to face the problem in 6D, where $[h_{\mu\nu}]=2$ and, consequently, the gauge parameter would have been given vanishing dimensions, as usual in gauge field theory. But the most general invariant action \eqref{Sinv} consists of two terms: $S_{LG}$ \eqref{SLG} and $S_{fract}$ \eqref{Sfract} which are respectively the actions for linearized gravity and for fractons, both physically relevant in 4D. Covariance makes evident that gravitons and fractons are indeed described by a unique gauge field theory, as already guessed in the non covariant approach. Precisely for this reason, it would be natural to have a unified theory where both contributions, gravitons and fractons, could be reached as limits of the complete theory. This legitimate expectation is not so easy to be satisfied, due, again, to the peculiarity of the defining symmetry \eqref{sym}. In fact, according to the standard Faddeev-Popov procedure, the gauge fixing condition should keep the same tensorial structure of the gauge parameter. In other words: to a scalar gauge parameter should correspond a scalar gauge condition, which in our case is \eqref{scalargaugecond},
which is the most general covariant one.
This has been done in \cite{Blasi:2022mbl}, where it has been shown that the theory exists only in the Landau gauge, since outside this particular gauge the quadratic gauge fixed action cannot be inverted and, consequently, the propagators do not exist. Being forced to work only in a specific gauge is not reassuring. First, because this does not happen in other more standard gauge field theories. Second, because there might be the concern that the obtained results are consequences of that gauge, hence unphysical and not really peculiar to the theory. 
Moreover, linearized gravity cannot be reached as a limit of the whole theory, for the obvious reason that it is defined by the infinitesimal diffeomorphism transformation \eqref{diff},
which is a gauge transformation depending on a vectorial gauge parameter, which by any means cannot be fixed by a scalar gauge condition, simply because four is greater than one. Evidence of this appears in the propagators, which display a singularity in the limit of vanishing fractonic contribution, which in this paper means $g_2\rightarrow 0$. The aim of this paper was to find a well defined gauge fixed theory, not constrained to a particular gauge choice, and where both limiting cases, fractonic and linearized gravity, could be reached smoothly. The idea is simply to look at the left hand side of \eqref{sym} and take the vectorial gauge fixing choice \eqref{vectorgaugecond},
which, again, is the most general covariant one and is the same of linearized gravity. It is not obvious at all that it could work. First, what does it mean that ``it works'' ? Besides the possibility of getting both fractons and gravitons as a limit, we asked two minimal requirements: that propagators are defined in a generic gauge, and that the counting of the degrees of freedom of the theory coincides with the one found in \cite{Blasi:2022mbl}, but without referring to a particular gauge choice. The result of this paper is that both requests have been achieved. We have now a covariant gauge fixed theory of fractons and linearized gravity, which has six degrees of freedom, or five, since for a particular combination of fractons and gravitons the theory is traceless. All the results are gauge independent, therefore the vector gauge fixing \eqref{vectorgaugecond} seems to be a better choice than the standard scalar one \eqref{scalargaugecond}.
We conclude this paper with a remark concerning the possible quantization of the action \eqref{Sinv}.
As it is well known, the quantization of LG is a long standing issue. As far as we know, a quantum field theory of fractons has not been achieved yet. In view of this, the covariant formulation adopted in our paper should be quite suitable, expecially because the fractonic part \eqref{Sfract} of the action \eqref{Sinv} impressively reminds the electromagnetic Maxwell theory. Hence, one might think about a kind of ``fracton QED'', where matter is coupled to fractons. Under this respect, the vector gauge fixing studied in this paper  can be very useful, not being restricted to the Landau gauge. 

\section*{Acknowledgments}\label{sec:acknowledgments}

We thank Stefano Giusto for enlightening discussions which motivated this work, which has been partially supported by the INFN Scientific Initiative GSS: ``Gauge Theory, Strings and Supergravity''. E.B. is supported by MIUR grant ``Dipartimenti di Eccellenza'' (100020-2018-SD-DIP-ECC\_001).
\appendix

\section{Basis for the $\Omega$-tensors}\label{sec:basis-for-the-omega-tensors}

In the momentum space gauge fixed action $S(g_1,g_2;\xi,\kappa)$ \eqref{gaugefixedaction} the kinetic operator $\tilde\Omega_{\mu\nu,\alpha\beta}(p)$ displays the following symmetries
\begin{equation}
		\tilde{\Omega}_{\mu\nu,\alpha\beta}(p) = \tilde{\Omega}_{\nu\mu,\alpha\beta}(p) = \tilde{\Omega}_{\mu\nu,\beta\alpha}(p) = \tilde{\Omega}_{\alpha\beta,\mu\nu}(p)\ .
\label{symmetries}	\end{equation}
It can be expanded on a basis formed by a set of five rank-4 tensors, collectively denoted $X_{\mu\nu,\alpha\beta}(p)$  \cite{Blasi:2015lrg,Gambuti:2020onb,Blasi:2017pkk,Gambuti:2021meo,Blasi:2022mbl,Amoretti:2013xya}
\be
X_{\mu\nu,\alpha\beta}\equiv (A,B,C,D,E)_{\mu\nu,\alpha\beta}
\label{baseX}\ee
with the same symmetry properties \eqref{symmetries}. Explicitly, the $X$-tensors read \cite{Blasi:2015lrg,Gambuti:2020onb,Blasi:2017pkk,Gambuti:2021meo,Blasi:2022mbl,Kugo:2014hja,Amoretti:2014iza}
\begin{align}
    A_{\mu \nu, \alpha \beta} &= \frac{d_{\mu \nu} d_{\alpha \beta}}{3} \label{A}\\[10pt]
 B_{\mu \nu, \alpha \beta} &= e_{\mu \nu} e_{\alpha \beta} \label{B}\\[10pt]
  C_{\mu \nu, \alpha \beta} &= \frac{1}{2} \left(  d_{\mu \alpha} d_{\nu \beta} + d_{\mu \beta} d_{\nu \alpha} - \frac{2}{3} d_{\mu \nu} d_{\alpha \beta}  \right) \label{C}\\[10pt]
  D_{\mu \nu, \alpha \beta} &=  \frac{1}{2} \left(  d_{\mu \alpha} e_{\nu \beta} + d_{\mu \beta} e_{\nu \alpha} + e_{\mu \alpha} d_{\nu \beta} + e_{\mu \beta} d_{\nu \alpha}  \right)\label{D}\\[10pt]
  E_{\mu \nu, \alpha \beta} &= \frac{\eta_{\mu \nu} \eta_{\alpha \beta}}{4} \label{E} \; ,
\end{align}
and $e_{\mu\nu}(p)$ and $d_{\mu \nu}(p)$ are the transverse and longitudinal projectors \eqref{projectors}, 
which are idempotent and orthogonal
\begin{equation}
    e_{\mu \lambda} {e^\lambda}_\nu = e_{\mu \nu}, \ \mathrm{}\   d_{\mu \lambda} {d^\lambda}_\nu = d_{\mu \nu}, \ \mathrm{}\     e_{\mu \lambda} {d ^\lambda}_\nu  =0 \: .  
\label{defed}\end{equation}
The $X$-tensors have the following properties:
\begin{itemize}
    \item 
    decomposition of the rank-4 tensor identity $\mathcal{I}_{\mu \nu, \alpha\beta}$ ~:
    \be
    A_{\mu \nu , \alpha \beta} + B_{\mu \nu , \alpha \beta} + C_{\mu \nu , \alpha \beta} + D_{\mu \nu , \alpha \beta} =  \mathcal{I}_{\mu \nu , \alpha \beta}
    \label{idempotency} 
    \ee
    \be
            \mathcal{I}_{\mu \nu, \rho \sigma} = \frac{1}{2} (\eta_{\mu \rho} \eta_{\nu \sigma} + \eta_{\mu \sigma} \eta_{\nu \rho}) 
\label{identity}\ee
    \item idempotency~:
    \begin{equation}
    X_{\mu\nu}^{\ \ \rho\sigma}X_{\rho\sigma,\alpha\beta}=X_{\mu\nu,\alpha\beta}\ ;
    \label{idempotency}\end{equation}
    \item orthogonality of $A$, $B$, $C$ and $D$~:
    \begin{equation}
    X_{\mu \nu , \alpha \beta} {{X^\prime}^{\alpha \beta}}_{\rho \sigma} = 0\ \ \mbox{if}\
    (X,X^\prime)\neq E\ \mbox{and}\ X\neq X^\prime\ ;
    \label{orthogonality}\end{equation}
    \item contractions with $E$~:
    \begin{align} 
    A_{\mu \nu , \alpha \beta} {E^{\alpha \beta}}_{\rho \sigma} 
    &= \frac{d_{\mu \nu} \eta_{\rho \sigma}}{4}\label{AE}\\[10pt] 
    B_{\mu \nu , \alpha \beta} {E^{\alpha \beta}}_{\rho \sigma} 
    &= \frac{e_{\mu \nu} \eta_{\rho \sigma}}{4}\label{BE}\\[10pt]
    C_{\mu \nu , \alpha \beta} {E^{\alpha \beta}}_{\rho \sigma}
    &=D_{\mu \nu , \alpha \beta} {E^{\alpha \beta}}_{\rho \sigma}=0\label{CE-DE}\ .    \end{align}
    
\end{itemize}

\section{Calculation of the propagators}\label{sec:calculation-of-the-propagators}

\subsection{$2g_1+g_2\neq0$}\label{sec:2g1g2neq0}

The matrix equation \eqref{eq:matriciale} yields 
\bea
			\tilde\Omega_{\mu\nu,\alpha\beta}\hat G^{\alpha\beta,\rho\sigma} + \tilde\Lambda^*_{\mu\nu,\lambda} \hat G^{*\lambda,\rho\sigma} &=& \mathcal{I}_{\mu\nu}^{\ \ \rho\sigma}\label{B1}\\
			\tilde\Omega_{\mu\nu,\alpha\beta}\hat G^{\alpha\beta,\tau} + \tilde\Lambda^*_{\mu\nu,\lambda}\hat G^{\lambda\tau} &=& 0\label{B2}\\
			\tilde\Lambda_{\gamma,\alpha\beta}\hat G^{\alpha\beta,\rho\sigma} + \tilde H_{\gamma\lambda}\hat G^{*\lambda,\rho\sigma} &=& 0\label{B3}\\
			\tilde\Lambda_{\gamma,\alpha\beta}\hat G^{\alpha\beta,\tau} + \tilde H_{\gamma\lambda}\hat G^{\lambda\tau} &=& \delta_{\gamma}^{\tau} \label{B4}\ .
\eea
The first equation \eqref{B1}, using the expansions \eqref{Omega}, \eqref{Lambda}, \eqref{hhprop} and the properties of the $X$-basis listed in Appendix \ref{sec:basis-for-the-omega-tensors}, gives a system of six equations (remember that our aim is to find the set of coefficients of the propagators \eqref{propcoeff})
\bea
			 4\tilde{t}\hat{t} + 3\tilde{t}\hat{w} - 4\tilde{v}\hat{v}  + 6p^2 \tilde{g}\hat{g} &=& 0\\
			 \tilde{t}\hat{w} + 2p^2 \tilde{g}\hat{l} &=&0\\
			 p^2 \tilde l\hat{l} &=& 2\\
			 p^2 \tilde l\hat{g} &=&0\\
			 \tilde{v}\hat{v} &=& 1\\
		\tilde{z}\hat{z}+p^2 \tilde{f}\hat{f} &=&1\ .
\eea
In the same way, from \eqref{B2}, \eqref{B3} and \eqref{B4} we get three more sets of equations
\bea
		2\tilde{z}\hat{f}+\tilde{f}\hat{r} &=&0\\
			 4\tilde{t}\hat{g} + 2\tilde{g}\hat{s} &=& 0\\
		\tilde{l}\hat{s} &=& 0\ ,
\eea

\bea
		 \tilde{f}\hat{z} + 2\tilde{r}\hat{f} &=&0\\
	4\tilde{g}\hat{t}+ 3\tilde{g}\hat{w}+\tilde{l}\hat{w}+8\tilde{s}\hat{g} &=&0\\
			4\tilde{l}\hat{u} +3\tilde{g}\hat{w}+\tilde{l}\hat{w}+8\tilde{s}\hat{l} &=&0\ ,
\eea
and
\bea
			p^2\tilde{f}\hat{f} + \tilde{r}\hat{r} &=&1\\
			3p^2 \tilde{g} \hat{g} + p^2 \tilde{l}\hat{l} +2\tilde{s}\hat{s} &=&2\ .
\eea
The above systems of equations are easily solved \cite{Bertolini:2020hgr} and, using the coefficients of the kinetic term \eqref{3.8} and \eqref{3.9}, we finally get
	\begin{align}
		&\hat{t} = \frac{(4\kappa+1)}{(\kappa+1)(2g_1+g_2)p^2}\label{eq:thatapp}\\*
		&\hat{u} = \frac{\kappa(4\kappa+1)-2\xi(2g_1+g_2)}{(\kappa+1)^2(2g_1+g_2)p^2}\label{eq:uhatapp}\\*
		&\hat{v} = \frac{1}{(g_2-g_1)p^2}\label{eq:vhatapp}\\*
		&\hat{z} = \frac{4\xi}{(2\xi g_2 -1)p^2}\label{eq:zhatapp}\\*
		&\hat{w} = \frac{-4\kappa}{(\kappa+1)(2g_1+g_2)p^2}\label{eq:whatapp}\\*
		&\hat{f} = \frac{-2}{(2\xi g_2-1)p^2}\label{eq:fhatapp}\\
		&\hat{g} = 0\label{eq:ghatapp}\\
		&\hat{l} = \frac{2}{(\kappa+1)p^2}\label{eq:lhatapp}\\
		&\hat{r} = \frac{4g_2}{(2\xi g_2-1)}\label{eq:rhatapp}\\
		&\hat{s} = 0\label{eq:shatapp}\ .
	\end{align}

\subsection{$2g_1+g_2=0$}\label{sec:2g1g20}

The action of the gauge fixed theory, after a field redefinition and setting $\kappa=0$ because this is the traceless case, is
\be
\left.S(g_1,g_2;\xi,\kappa)\right|_{2g_1+g_2=0;\kappa=0} = 
\left.S_{inv}(g_1,g_2)\right|_{2g_1+g_2=0} + S_{gf}(\xi)\ ,
\label{}\ee
where the invariant action $\left.S_{inv}(g_1,g_2)\right|_{2g_1+g_2=0}$ and  the gauge fixing term $S_{gf}(\xi)$ are given by \eqref{2g1+g_2} and \eqref{Sgftraceless}, respectively \cite{Bertolini:2021iku}. It can be written in the form \eqref{momentumaction}, and the coefficients in $\tilde\Omega_{\mu\nu,\alpha\beta}$ \eqref{Omega}, $\tilde\Lambda_{\alpha\beta,\mu}$ \eqref{Lambda} and $\tilde H_{\mu\alpha}$ \eqref{H} are 
	\begin{align}
		\tilde{t} &= -3p^2 &
		\tilde{u} &= p^2 &
		\tilde{v} &= -3p^2&
		\tilde{z} &= -p^2;&
		\tilde{w} &= 0 \label{B30}\\
		\tilde{f} &= \frac{1}{2}&
		\tilde{g} &= 0&
		\tilde{l} &= 1&
		\tilde{r} &= \frac{\xi}{2}&
		\tilde{s} &= \frac{\xi}{2}\ .
	\end{align}

Following the same steps of the general case, we find the system of equations for the coefficients of the propagators
	{\small \bea
			\tilde{v}\hat{v} &= 1\label{eq:systemea1}\\
			\tilde{z}\hat{z}+p^2 \tilde{f}\hat{f} &= 1\label{eq:systemea2}\\
			p^2\tilde{f}\hat{f} + \tilde{r}\hat{r} &= 1\label{eq:systemea3}\\
			2\tilde{z}\hat{f}+\tilde{f}\hat{r} &= 0\label{eq:systemea4}\\
			\tilde{f}\hat{z} + 2\tilde{r}\hat{f} &= 0\label{eq:systemea5}\\
			p^2 \hat{l} +2\tilde{s}\hat{s} &= 2\label{eq:systemea6}\\
			4\tilde{t}\hat{g} &= 0\label{eq:systemea7}\\
			4\tilde{u}\hat{l} + 2\hat{s} &= 0\label{eq:systemea8}\\
			4\tilde{t}\hat{t} + 3\tilde{t}\hat{w} &= 4\label{eq:systemea9}\\
			\hat{w}+8\tilde{s}\hat{g} &= 0\label{eq:systemea10}\\
			\tilde{u}\hat{w} + 2p^2 \hat{g} &= 0\label{eq:systemea11}\\
			4\tilde{u}\hat{u} + \tilde{u}\hat{w} +2p^2 \hat{l} &= 4\label{eq:systemea12}\\
			\tilde{t}\hat{w} &= 0\label{eq:systemea13}\\
			4\hat{u}+\hat{w}+8\tilde{s}\hat{l} &= 0\ ,\label{eq:systemea14}
	\eea}
	\hspace{-4pt}
which is simpler than the general case. In particular, since $\tilde t(p)\neq0$ in \eqref{B30}, from \eqref{eq:systemea7} and \eqref{eq:systemea9} we immediately get $\hat g = \hat w =0$. 
The solutions are therefore easily found 
	{\footnotesize \begin{align}
			\hat{t} &= -\frac{1}{3p^2}&
			\hat{u} &= \frac{2\xi}{(2\xi-1)p^2}&
			\hat{v} &= -\frac{1}{3p^2}&
			\hat{z} &= \frac{-4\xi}{(4\xi+1)p^2}&
			\hat{w} &= 0\\
			\hat{f} &= \frac{2}{(4\xi+1)p^2}&
			\hat{g} &= 0&
			\hat{l} &= \frac{-2}{(2\xi-1)p^2}&
			\hat{r} &= \frac{8}{(4\xi+1)}&
			\hat{s} &= \frac{4}{(2\xi-1)}\ . 
	\end{align}}

\end{document}